\documentclass[aps,prl,reprint,twocolumn]{revtex4}
\usepackage{blindtext}
\usepackage[T1]{fontenc}
\usepackage{amsmath}
\usepackage{amssymb}
\usepackage{xcolor}

\definecolor{green}{rgb}{0,0,1}

\makeatletter

\usepackage[utf8]{inputenc}

\usepackage{amsmath}
\usepackage{amsfonts}
\usepackage{graphicx}
\usepackage{yfonts}
\usepackage{color}
\usepackage[normalem]{ulem}
\usepackage{amsthm}
\usepackage{bm}
\usepackage{bbm}
\usepackage{mathtools}
\usepackage{array}
\usepackage{placeins}
\usepackage{enumitem}

\def\<{\langle}
\def\>{\rangle}

\newcommand{\Tr}{\mathrm{Tr}}

\DeclareMathAlphabet\mathbfcal{OMS}{cmsy}{b}{n}

\mathchardef\mhyphen="2D 

\newtheorem{Theorem}{Theorem}
\newtheorem{Lemma}{Lemma}

\newtheorem{Definition}{Definition}

\usepackage{babel}

\makeatother

\usepackage{babel}
\begin{document}

\title{Violation of steering inequality  for generalized equiangular measurements}

\author{Adam Rutkowski${}^1$\footnote{e-mail:adam.rutkowski@ug.edu.pl} and Katarzyna Siudzi\'{n}ska${}^2$}

\affiliation{ ${}^1$Institute of Theoretical Physics and Astrophysics \\ Faculty of Mathematics, Physics and Informatics, University of Gda\'{n}sk, 80-952 Gda\'{n}sk, Poland\\
${}^2$Institute of Physics, Faculty of Physics, Astronomy and Informatics,\\ Nicolaus Copernicus University in Toru\'{n}, ul. Grudzi\c{a}dzka 5, 87-100 Toru\'{n}, Poland}


\begin{abstract}
In this article, we study bipartite quantum steering using a general class of measurement operators known as the generalized equiangular measurement (GEAM). Our approach allows for the construction of steering inequalities that are applicable not only to informationally complete measurements but also to incomplete or lossy scenarios, where the resulting quantum assemblages may be subnormalized. 
We develop a method to analytically bound the value of steering functionals under local hidden state (LHS) models by employing a generalized inequality for the eigenvalues of positive semidefinite operators, extending classical results such as the Samuelson inequality. This yields universal upper bounds on the LHS value in terms of the parameters defining the frame.
Furthermore, we analyze the asymptotic behavior of these bounds in high-dimensional Hilbert spaces and demonstrate the possibility of \emph{unbounded quantum violations} of steering inequalities. Our formalism encompasses previously known results for mutually unbiased bases (MUBs) and mutually unbiased measurements (MUMs) as special cases, showing that these are particular instances within a unified and more general framework.
\end{abstract}



\maketitle


\section{Introduction}

This article delves into the study of quantum steering, a phenomenon in quantum mechanics that illustrates nonlocal correlations between two observers. Introduced by Schrodinger in 1935 \cite{schrod}, quantum steering extended the ideas of the EPR paradox \cite{EPR} to bipartite systems, where one party's measurements can influence the state of another party's system through entanglement. This interaction allows one observer to ``steer'' the quantum state of the other, demonstrating an intriguing form of quantum correlation.

Quantum steering is a distinct type of nonlocality, situated between general entanglement and Bell nonlocality \cite{wieseman2007}. While all steerable states are entangled, not all entangled states exhibit steering, and some steerable states do not violate Bell inequalities \cite{wieseman2007,Quintino2015} . Experimental confirmation of steering relies on steering inequalities \cite{Cavalcanti}, though relatively few such inequalities have been constructed to date. Several works \cite{Cavalcanti, WSGTR2011} use uncertainty relations to derive steering inequalities for continuous or specific discrete measurements.

In this work, we propose a new method to derive steering inequalities using \emph{generalized equiangular measurements} (GEAMs) \cite{GEAM}, a broad class of operator systems that generalize mutually unbiased measurements {\cite{Kalev} and generalized symmetric, informationally complete (SIC) measurements \cite{Gour}}. These constructions allow for greater flexibility and structural diversity in measurement design. Notably, our framework encompasses both informationally complete and incomplete measurements, and thus naturally models assemblages that are \emph{subnormalized}, i.e., those for which the total trace  may be strictly less than one. This generalization reflects realistic physical scenarios such as lossy detection, partial measurement access, or coarse-graining, and is consistent with the nonsignaling principle.

To analyze the performance of steering functionals built from GEAMs, we derive explicit bounds for local hidden state (LHS) models using a spectral method. In particular, we utilize an inequality bounding the extreme eigenvalues of positive semidefinite operators, a generalized version of Samuelson's inequality, known in matrix analysis and operator theory \cite{Wolkowicz1980, bhatia_matrix_1997, Sharma1994}. This approach allows us to obtain universal upper bounds on the LHS value for any such steering functional, expressed analytically in terms of {measurement} parameters.

We then study the asymptotic behavior of these bounds in high-dimensional Hilbert spaces. Remarkably, we show that in certain regimes, the quantum violation $V(F)$ defined as the ratio of the quantum value to the LHS bound becomes \emph{unbounded}, growing arbitrarily large with dimension. This result demonstrates the fundamental power of generalized measurements in producing extreme nonclassical correlations.

Interestingly, several known results from the literature emerge in our framework as special cases corresponding to particular choices of the measurement operators. In particular, we recover the previously derived bounds for mutually unbiased bases (MUBs) and mutually unbiased measurements (MUMs) by specializing the parameters of {the} generalized equiangular measurements. This demonstrates that our formulation not only generalizes prior work but also unifies disparate constructions under a common spectral framework.
Violation can become \emph{unbounded}, revealing new insights into the fundamental structure of steering and the power of generalized measurements.

\section{Preliminaries}

{In this section, we introduce the mathematical framework for steering inequalities and the structure of assemblages. We begin by defining quantum assemblages and local hidden state (LHS) models. Next, we describe the steering functionals constructed from generalized equiangular measurements (GEAMs).}

\subsection*{Quantum assemblages}

An \emph{assemblage} is a collection of subnormalized quantum states $\mathcal{Q} = \{\sigma_{\alpha,k}\}$, resulting from measurements performed on one part of an entangled bipartite system.

\begin{Definition}[Standard assemblage]
\label{d:assem}
An assemblage is a collection $\mathcal{Q} = \{\sigma_{\alpha,k} :\, \alpha = 1, \ldots, L;\; k = 1, \ldots, M_\alpha\}$ of $d \times d$ Hermitian matrices (positive semidefinite operators) such that:
\begin{enumerate}

    \item $\sigma_{\alpha,k} \geq 0$ for all $\alpha, k$;
    \item {$\mathrm{Tr} \left( \sum_{k=1}^{M_\alpha} \sigma_{\alpha,k} \right) = 1$} for all $\alpha$.
\end{enumerate}
\end{Definition}
It is well known \cite{Sch1936,HJW1993} that such assemblages can be realized by performing positive operator-valued (POVM) measurements on a bipartite quantum state. Let $\mathcal{H}_A$ and $\mathcal{H}_B$ denote the Hilbert spaces corresponding to Alice and Bob, respectively, and let $\rho \in B(\mathcal{H}_A \otimes \mathcal{H}_B)$ be a bipartite quantum state shared between the two parties.
Specifically, for each measurement setting $\alpha$, Alice performs a measurement  $\{P_{\alpha,k}\}$ on her subsystem. Here, $P_{\alpha,k} \in B(\mathcal{H}_A)$ denotes the measurement operator associated with outcome $k$ of setting $\alpha$. The resulting (subnormalized) quantum states on Bob’s side—also called assemblage elements—are then given by:
\begin{equation}
\sigma_{\alpha,k} = \mathrm{Tr}_A \left[ (P_{\alpha,k} \otimes \mathbb{I}_B) \rho \right],
\end{equation}
where $\mathrm{Tr}_A$ denotes the partial trace over Alice’s subsystem. The collection $\{\sigma_{\alpha,k}\}$ forms the quantum assemblage observed by Bob.

\paragraph{Subnormalized assemblages.} In this work, we generalize Definition~\ref{d:assem} by allowing \emph{subnormalized} assemblages, in which the trace condition is relaxed:
\begin{equation}
\mathrm{Tr} \left( \sum_{k=1}^{M_\alpha} \sigma_{\alpha,k} \right) \leq 1 \quad \text{for all } \alpha.
\end{equation}
This generalization allows us to model realistic or restricted experimental scenarios, including: partial or lossy measurements (e.g., detection inefficiencies)~\cite{detection}, coarse-grained measurements or partial outcome sets~\cite{coarse,Lukasz}, measurements that are informationally incomplete~\cite{EOM24}, and post-selection on subsets of outcomes~\cite{post}.

Even in this generalized setting, the \emph{nonsignaling condition} may still be satisfied. It requires that Bob's reduced state is independent of Alice’s measurement setting~$\alpha$, which corresponds to the \emph{normalization constraint} in Definition~\ref{d:assem}, item~(2). That is, for any choice of~$\alpha$, the sum of the conditional states $\sum_k \sigma_{\alpha,k}$ must yield the same marginal state on Bob’s side. This ensures that Alice’s choice of measurement cannot be inferred from Bob’s local statistics, in accordance with the relativistic no-signaling principle~\cite{nonsignaling-review}.

\subsection*{Local hidden state models}

In the context of quantum steering, local hidden state (LHS) models serve as the analogue of local hidden variable (LHV) models familiar from Bell nonlocality. While LHV models explain correlations between measurement outcomes using shared classical randomness and pre-existing values, LHS models go a step further: they allow only Bob's system to be described by quantum states, while Alice's outcomes are governed by hidden variables.

This asymmetry reflects the steering scenario, where only Bob's device is trusted, and Alice's measurements are uncharacterized. LHS models thus capture the idea that any apparent steering observed in experiments could, in principle, be simulated using shared randomness and a local quantum system on Bob's side.

\begin{Definition}[Local hidden state (LHS) model]
\label{def:LHS}
An assemblage $\mathcal{Q} = \{\sigma_{\alpha,k}\}$ is said to admit a \emph{local hidden state (LHS)} model if there exist:
\begin{itemize}
    \item a finite index set $\Lambda$,
    \item a probability distribution $\{q_\lambda\}_{\lambda \in \Lambda}$,
    \item a collection of normalized quantum states $\{\sigma_\lambda\}_{\lambda \in \Lambda} \subset B(\mathcal{H}_B)$,
    \item and conditional probability distributions $p_\lambda(k|\alpha)$,
\end{itemize}
such that, for all measurement settings $\alpha$ and outcomes $k$, the elements of the assemblage can be written as:
\begin{equation}
\sigma_{\alpha,k} = \sum_{\lambda \in \Lambda} q_\lambda \, p_\lambda(k|\alpha) \, \sigma_\lambda.
\end{equation}
\end{Definition}

The set of all assemblages that admit such a model is denoted by $\mathcal{L}$. Assemblages outside $\mathcal{L}$ are referred to as steerable.

\begin{figure}\centering
\includegraphics[width=8cm]{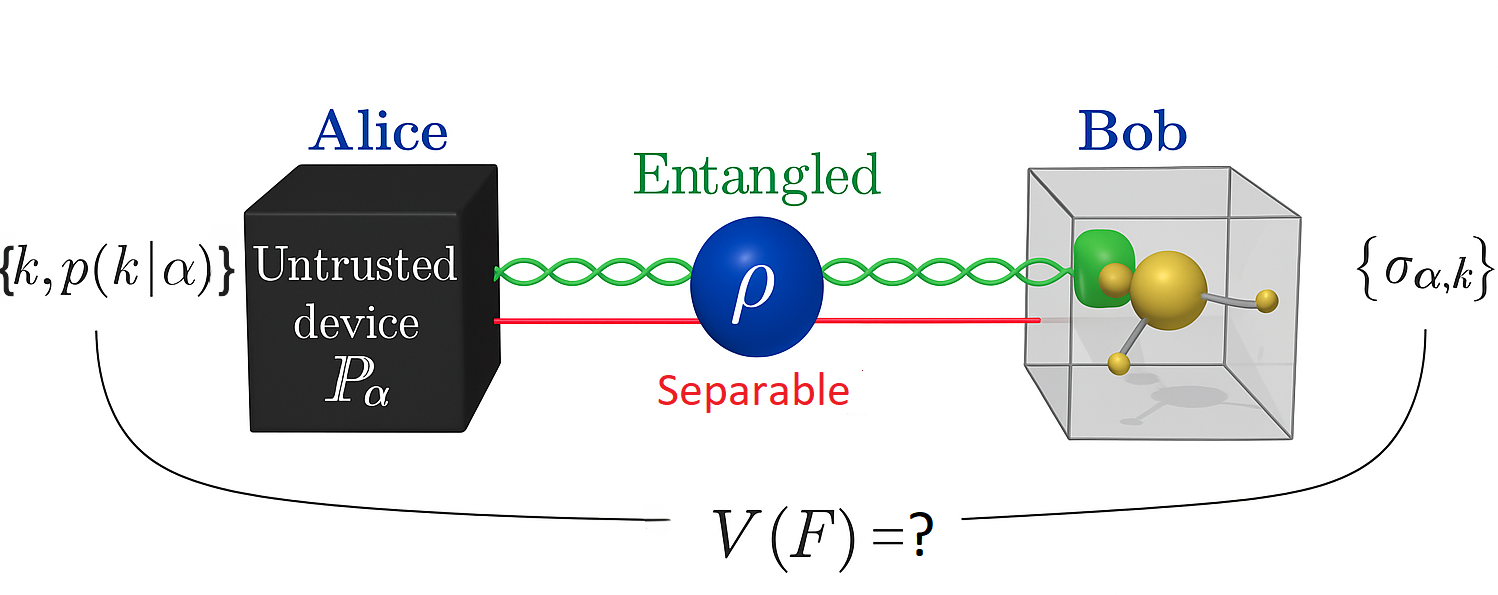}
\caption{Quantum steering scenario: Alice and Bob share a bipartite state $\rho$, either entangled (green) or separable (red). Alice performs a measurement labeled by $\alpha$ using an untrusted device (a black box), with measurement outcomes $k$ occurring with probability $p(k|\alpha)$. She communicates $k$ to Bob, who receives the corresponding conditional state $\sigma_{\alpha,k}$ and analyzes the resulting assemblage. Bob then computes the value of a steering functional and checks for violation of a steering inequality --- see Definitions~\ref{def:steeringineq} and~\ref{def:bound}.}
\label{fig:scheme}
\end{figure}

\subsection*{Steering inequalities and functionals}

To test whether a given assemblage exhibits quantum steering, we evaluate it using a linear functional constructed from a family of Hermitian operators.

\begin{Definition}[Steering functional]
\label{def:steeringineq}
Let $L$ and $M_\alpha$ be positive integers, and let $d$ be the Hilbert space dimension. A \emph{steering functional} is a collection $F = \{F_{\alpha,k}\}$ of $d \times d$ Hermitian matrices, indexed by measurement settings $\alpha = 1, \ldots, L$ and outcomes $k = 1, \ldots, M_\alpha$. Its value on an assemblage $\sigma = \{\sigma_{\alpha,k}\}$ is defined as:
\begin{equation}
\langle F, \sigma \rangle := \mathrm{Tr} \left( \sum_{\alpha=1}^L \sum_{k=1}^{M_\alpha} F_{\alpha,k} \, \sigma_{\alpha,k} \right).
\end{equation}
\end{Definition}

This evaluation yields a real number associated with the assemblage and the chosen steering functional.

\begin{Definition}[Steering inequality and quantum violation]
\label{def:bound}
Let $F$ be a steering functional. We define the following quantities:
\begin{itemize}
    \item The \emph{LHS bound}:
    \begin{equation}
    S_{\mathrm{LHS}}(F) := \sup_{\sigma \in \mathcal{L}} |\langle F, \sigma \rangle|,
    \end{equation}
    \item The \emph{quantum bound}:
    \begin{equation}
    S_Q(F) := \sup_{\sigma \in \mathcal{Q}} |\langle F, \sigma \rangle|,
    \end{equation}
    \item The \emph{quantum violation ratio}:
    \begin{equation}
    V(F) := \frac{S_Q(F)}{S_{\mathrm{LHS}}(F)}.
    \end{equation}
\end{itemize}
A \emph{steering inequality} is then the statement:
\begin{equation}
|\langle F, \sigma \rangle| \leq S_{\mathrm{LHS}}(F),
\end{equation}
which holds for all assemblages $\sigma \in \mathcal{L}$ that admit an LHS model. If $V(F) > 1$, the inequality is violated by some quantum assemblage, certifying the presence of steering.
\end{Definition}

\subsection*{Generalized equiangular measurements}

{Recently, a wide class of informationally overcomplete measurements has been introduced. It is based in equiangularity between operators \cite{EAL3} and generalizes the notion of equiangular tight frames \cite{ETF1}.}

\begin{Definition}[Generalized equiangular measurement \cite{GEAM}]
\label{geam}
A collection of $N$ sets of operators $\{P_{\alpha,k}\}_{k=1}^{M_\alpha}$ for $\alpha = 1, \ldots, N$ is a \emph{generalized equiangular measurement} (GEAM) if:
\begin{equation}
\sum_{k=1}^{M_\alpha} P_{\alpha,k} = \gamma_\alpha \mathbb{I}_d, \qquad \sum_{\alpha=1}^N \gamma_\alpha = 1,
\end{equation}
and the following relations hold:
\begin{align}
\mathrm{Tr}(P_{\alpha,k}) &= a_\alpha, \\
\mathrm{Tr}(P_{\alpha,k}^2) &= b_\alpha \, a_\alpha^2, \\
\mathrm{Tr}(P_{\alpha,k} P_{\alpha,\ell}) &= c_\alpha \, a_\alpha^2, \quad (k \neq \ell), \\
\mathrm{Tr}(P_{\alpha,k} P_{\beta,\ell}) &= f \, a_\alpha a_\beta, \quad (\alpha \neq \beta).
\end{align}
\end{Definition}

{The measurement parameters} are determined as:
\begin{equation}
a_\alpha = \frac{d \gamma_\alpha}{M_\alpha}, \qquad
c_\alpha = \frac{M_\alpha - d b_\alpha}{d(M_\alpha - 1)}, \qquad
f = \frac{1}{d},
\end{equation}
{and the admissible} values of $b_\alpha$ satisfy:
\begin{equation}\label{ba}
\frac{1}{d} < b_\alpha \leq \frac{1}{d} \min\{d, M_\alpha\}.
\end{equation}
{Moreover, the total number of elements
\begin{equation}
\sum_{\alpha=1}^NM_\alpha=d^2+N-1.
\end{equation}
Note that the operators $P_{\alpha,k}$ form an informationally overcomplete set for any $N\geq 2$, whereas $N=1$ corresponds to generalized symmetric, informationally complete (SIC) POVMs \cite{Gour}. However, one can always choose $L<N$ sets to form an informationally incomplete collection. In this case, the corresponding assemblages may be subnormalized. This forms the foundation for our generalized steering analysis.}

\section{Violation of steering inequality}

In this section, we demonstrate that steering inequalities based on generalized equiangular measurements (GEAMs) can lead to unbounded quantum violations as the dimension of the Hilbert space grows. We provide an explicit analytical upper bound for the LHS value, show its asymptotic decay with dimension, and compare it with the quantum value.

\subsection*{LHS bound via spectral inequality}

Let $F = \{P_{\alpha,k}\}$ be a steering functional constructed from $L \leq N$ measurement settings of a generalized equiangular measurement, with $\alpha = 1,\ldots,L$ and $k = 1,\ldots,M_\alpha$. We consider the action of $F$ on an LHS assemblage of the form:
\begin{equation}
\sigma_{\alpha,k} = \sum_{\lambda \in \Lambda} q_\lambda\, p_\lambda(k|\alpha)\, \sigma_\lambda,
\end{equation}
with $\{q_\lambda\}$ being a probability distribution and $\{\sigma_\lambda\}$ a set of fixed quantum states. The value of the functional becomes:
\begin{equation}\label{eq:lhs-eval}
\langle F, \sigma \rangle = \sum_{\alpha,k} \mathrm{Tr}\left(P_{\alpha,k} \sigma_{\alpha,k}\right)
= \sum_{\lambda} q_\lambda \, \mathrm{Tr}\left(Y_\lambda \sigma_\lambda\right),
\end{equation}
where we define the  operator
\begin{equation}\label{eq:Ylambda}
Y_\lambda := \sum_{\alpha=1}^L \sum_{k=1}^{M_\alpha} p_\lambda(k|\alpha) P_{\alpha,k}.
\end{equation}
Using the operator norm $\|Y_\lambda\| = \eta_{\max}(Y_\lambda)$ that is equal to the maximal eigenvalue $\eta_{\max}$ of $Y_\lambda$, we obtain
\begin{equation}
\langle F, \sigma \rangle \leq \sup_\lambda \|Y_\lambda\|.
\end{equation}
 This follows from the duality between \( \ell^1(\Omega; \mathcal{S}_1^d) \) and \( \ell^\infty(\Omega; \mathcal{M}_d) \), as detailed in the Supplemental Material of \cite{Rutkowski2017}.

To estimate this norm, we employ a spectral inequality for positive semidefinite operators (a generalization of the Samuelson inequality), bounding the largest eigenvalue as:
\begin{equation}\label{eq:xi_bound}
\eta_{\max}(Y_\lambda) \leq \xi_+ := \frac{1}{d} \left[ \sum_{\alpha=1}^L a_\alpha + \sqrt{(d-1) \sum_{\alpha=1}^L a_\alpha^2 (d b_\alpha - 1)} \right],
\end{equation}
where $a_\alpha = d\gamma_\alpha / M_\alpha$ and $b_\alpha$ are the parameters of the GEAM (see Section~\ref{geam} for definitions). The full derivation of this bound is presented in \textbf{Appendix A}.

Thus, we obtain the analytical LHS bound:
\begin{equation}
S_{\mathrm{LHS}} \leq \xi_+.
\end{equation}

\subsection*{Asymptotic behavior and violation}

We now examine the behavior of $\xi_+$ in the large-$d$ limit. Assuming the following parametrization of $b_\alpha$,
\[
b_\alpha = \frac{z_\alpha}{d}, \quad 1<z_\alpha\leq\min\{d,M_\alpha\},
\]
we obtain
\begin{equation}
\xi_+=\sum_{\alpha=1}^L\frac{\gamma_\alpha}{M_\alpha}+
    \sqrt{\Delta(d)},
\end{equation}
where
\[
\Delta(d) := (d-1)\sum_{\alpha=1}^L\frac{\gamma_\alpha^2}{M_\alpha^2}(z_\alpha-1).
\]
If $\Delta(d) = \mathcal{O}(1/d^p)$ with $p>0$, then $\xi_+<1$ for large dimensions $d$. A suitable choice of the measurement parameters is detailed in {\bf{Appendix B}}.

We now state the main result of this section.

\begin{Theorem}[Violation of steering inequality from GEAMs]
\label{thm:violation}
Let $F$ be a steering functional constructed from $L$ measurement settings of a generalized equiangular measurement (GEAM). Then, the corresponding LHS bound $\xi_+$ is a decreasing function of the Hilbert space dimension $d$ and satisfies
\begin{equation}
\xi_+ < 1 \quad \text{for sufficiently large } d.
\end{equation}
Consequently, the steering inequality
\begin{equation}
\langle F, \sigma \rangle \leq \xi_+
\end{equation}
and the resulting quantum violation
\begin{equation}
V(F) := \frac{S_Q}{S_{\mathrm{LHS}}} \geq \frac{1}{\xi_+}
\end{equation}
grows without bound as $d \to \infty$.
\end{Theorem}

\paragraph{Remark.} The quantum value $S_Q$ is calculated explicitly in \textbf{Appendix B} for a representative class of quantum assemblages. It is shown that $S_Q = 1$ under appropriate normalization, confirming that the violation ratio indeed diverges in the large-$d$ limit.


\section{Special cases}

{\subsection*{Mutually unbiased bases}}

{Mutually unbiased bases (MUBs) are represented by a collection of $L\leq d+1$ sets of $d$ orthogonal rank-1 projectors, where the transition probability between any two operators that belong to different sets is constant \cite{Durt,MAX}. They correspond to the following choice of parameters:
    \begin{equation}
        a_\alpha=\frac 1L,\qquad b_\alpha=1,
    \end{equation}
for which
    \begin{equation}
    \xi_+=\frac 1d \left(1+\frac{d-1}{\sqrt{L}}\right),
    \end{equation}
agrees with the result from ref. \cite{PRL-steering}. For infinite steering violation, the number of bases $L$ has to depend on the dimension $d$. Otherwise, $\xi_+\to 1/\sqrt{L}$ as $d\to\infty$. In particular, $L=3$ (the number of MUBs known to exist in any dimension) results in $\xi_+=1/\sqrt{3}$.}

{\subsection*{Mutually unbiased measurements}}

{Mutually unbiased measurements (MUMs) generalize the notion of mutually unbiased bases to nonprojective measurements characterized via the parameter $\kappa$ \cite{Kalev}. Contrary to the MUB case, MUMs exist in every finite dimension. A set of $L\leq d+1$ MUMs is recovered for
    \begin{equation}
        a_\alpha=\frac 1L,\qquad b_\alpha=\kappa,\qquad \frac 1d <\kappa\leq 1.
    \end{equation}
In this case, eq. (\ref{xip}) reduces to
    \begin{equation}
    \xi_+=\frac 1d \left(1+\sqrt{\frac{(d-1)(d\kappa-1)}{L}}\right).
    \end{equation}
Observe that the introduction of a free parameter $\kappa$ allows for infinite steering violation under a weaker condition
\begin{equation}
\lim_{d\to\infty}\frac{\kappa}{L}=0.
\end{equation}
This holds e.g. for a constant $\kappa$ and $L$ being a linear function of $d$, or $\kappa$ being a linear function of $1/d$ and a constant $L$.}

{\subsection*{Symmetric measurements}}

\begin{figure}\centering
\includegraphics[width=8cm]{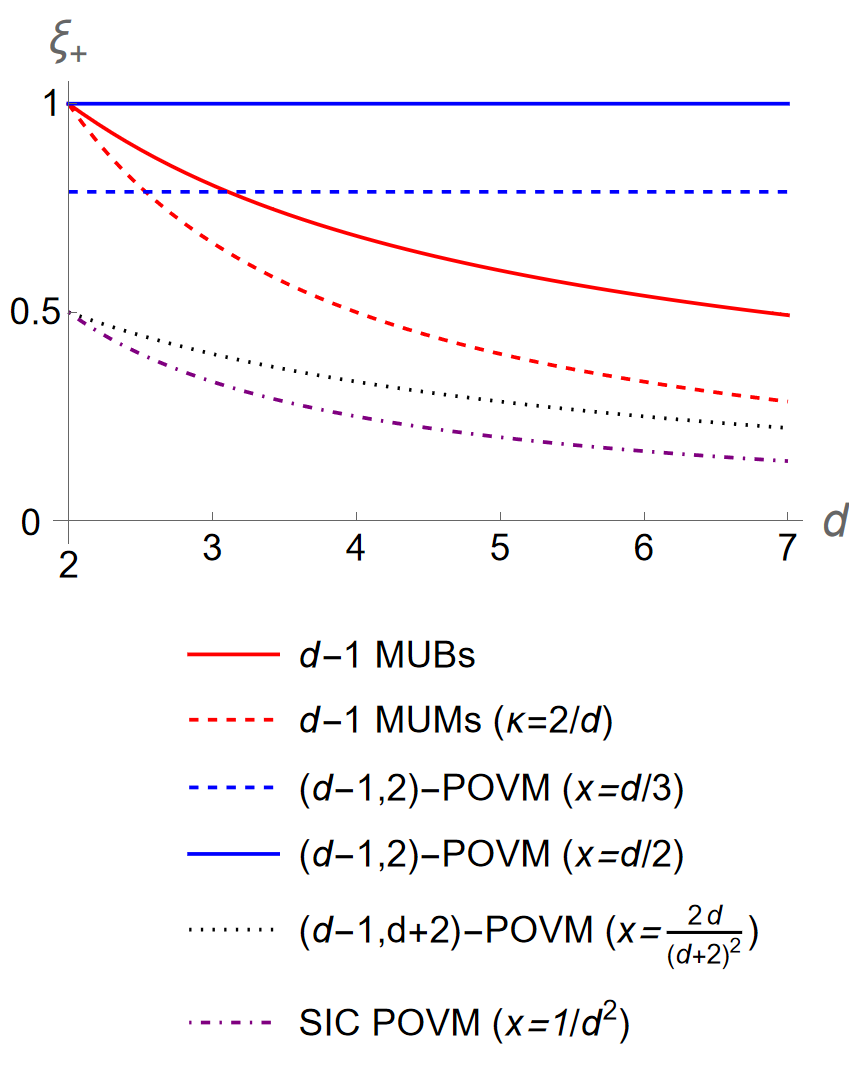}
\caption{The quantum violation $V(F)=\xi_+$ as a function of dimension $d$ depending on the choice of measurement operators.}
\label{fig:plot}
\end{figure}

{A class of symmetric measurements, also known as $(N,M)$-POVMs, can be considered as a generalization of both mutually unbiased bases and SIC POVMs \cite{SIC-MUB}. An informationally incomplete $(L,M)$-POVM follows from $M_\alpha=M$ and
    \begin{equation}
        a_\alpha=\frac{d}{ML},\qquad b_\alpha=\frac{xM^2}{d^2},
    \end{equation}
    \begin{equation}
        \frac{d}{M^2}<x\leq\min\left\{\frac{d^2}{M^2},\frac {d}{M}\right\}.
    \end{equation}
Now, one has
    \begin{equation}
    \xi_+=\frac 1M \left(1+\sqrt{\frac{(d-1)(xM^2-d)}{dL}}\right),
    \end{equation}
and hence $\xi_+\to 0$ as $d\to\infty$ as long as $M$ is a linear function of $d$ and
\begin{equation}
\lim_{d\to\infty}\frac{x}{L}=0.
\end{equation}
An example of an $(L,M)$-POVM that goes beyond generalizations of MUBs and SIC POVMs corresponds to the choice $L=d-1$ and $M=d+2$, for which $\xi_+$ is plotted in Fig. \ref{fig:plot}.
}

Another special case belongs to dichotomic measurements, which are measurements with only two possible outcomes ($M=2$). This time, however, $\xi_+$ always has a finite, non-zero value.

{\subsection*{Projective measurements}}

{
    If the generalized equiangular measurements are constructed from equiangular tight frames, then they are projective measurements ($P_{\alpha,k}$ are of rank-1). Assume that each frame has $M$ elements. Then, the parameters
    \begin{equation}
        a_\alpha=\frac{d\gamma_\alpha}{M},\qquad b_\alpha=1,
    \end{equation}
    and hence
    \begin{equation}
    \xi_+=\frac 1M \left[1+(d-1)\sqrt{\sum_{\alpha=1}^L\gamma_\alpha^2}\right].
    \end{equation}
    In this example, $\xi_+\to 0$ with $d\to\infty$ if $M$ is a linear function of $d$ and
    \begin{equation}
    \lim_{d\to\infty}\sum_{\alpha=1}^L\gamma_\alpha^2=0.
    \end{equation}
In particular, for SIC POVMs ($M=d^2$, $N=1$, and $\gamma_\alpha=1$), one has simply
    \begin{equation}
    \xi_+=\frac{1}{d}.
    \end{equation}
}

\section{Conclusions and discussion}

In this work, we have presented a general framework for analyzing quantum steering using generalized equiangular measurements (GEAMs). By allowing for subnormalized assemblages that arise from incomplete or lossy measurements, we extended the conventional formalism beyond the standard informationally complete settings. This generalization is not only physically motivated but also mathematically tractable, as shown by the consistent satisfaction of the nonsignaling condition.
We constructed a family of steering inequalities based on GEAMs and derived analytical upper bounds on their local hidden state (LHS) values. These bounds are expressed in terms of the {measurement} parameters and rely on a spectral inequality for the maximal eigenvalue of positive semidefinite operators.

Our main result is the demonstration that, under certain choices of GEAM parameters and in the asymptotic limit of high-dimensional Hilbert spaces, the steering violation can become \textit{unbounded}. That is, the ratio between the quantum value and the LHS bound of the steering functional can grow without limit as the dimension increases. This contrasts with previous results \cite{Rutkowski2017,PRL-steering}, which typically yielded bounded violations for specific  classes of measurements.
Importantly, our framework recovers known steering bounds for mutually unbiased bases (MUBs) and mutually unbiased measurements (MUMs) as particular cases. This unification confirms the robustness and generality of our method. The resulting inequalities and bounds provide not only deeper insights into the structure of quantum correlations but also practical tools for experimental and theoretical investigations of steering in generalized measurement scenarios.

Several open {questions} remain. On the theoretical side, it would be interesting to investigate whether our bounds are tight in the general GEAM case or whether stronger violations may arise with optimized quantum states. On the experimental side, implementing {informationally incomplete subsets of GEAMs} in optical or atomic platforms could provide novel demonstrations of quantum steering beyond conventional designs.

\section{Acknowledgements}

This research was funded in whole or in part by the National Science Centre, Poland, Grant number 2021/43/D/ST2/00102. For the purpose of Open Access, the author has applied a CC-BY public copyright licence to any Author Accepted Manuscript (AAM) version arising from this submission.

\section*{Appendix A: Bounds on the largest eigenvalue of $Y_\lambda$}

To prove the upper bound of $\<F,\sigma\>$, we apply the following results.

\begin{Lemma}(Generalized Samuelson's inequality \cite{Sharma1994})
    The eigenvalues $\eta_k$ of a positive-semidefinite $d\times d$ matrix $X$ are bounded by
    \begin{equation}\label{bounds}
        m-s\frac{d-1}{\sqrt{d}}\leq\eta_k\leq m+s\frac{d-1}{\sqrt{d}},
    \end{equation}
    where
    \begin{equation}
        m=\frac 1d \Tr (X),\qquad s=\sqrt{\frac{1}{d-1}(\Tr X^2-dm^2)}.
    \end{equation}
    The upper and lower bounds are reached if and only if $\eta_k\geq\eta$ or $\eta_k\leq\eta$, respectively, where $\eta_j=\eta$ for all $j\neq k$.
\end{Lemma}

For $Y_\lambda$ in eq. (\ref{eq:Ylambda}), we calculate
    \[
    \begin{split}
        \Tr (Y_\lambda) &= \Tr\left(\sum_{\alpha=1}^L\sum_{k=1}^{M_\alpha}p_\lambda(k|\alpha)P_{\alpha,k}\right)
    \\&=\sum_{\alpha=1}^L\sum_{k=1}^{M_\alpha}p_\lambda(k|\alpha) a_\alpha
        =\sum_{\alpha=1}^La_\alpha\equiv A
    \end{split}
    \]
and
    \[
    \begin{split}
    \Tr(Y_\lambda^2)
    =&\Tr\left(\sum_{\alpha,\beta=1}^L\sum_{k,\ell=1}^{M_\alpha}
        p_\lambda(k|\alpha)p_\lambda(\ell|\beta)P_{\alpha,k}P_{\beta,\ell}\right)\\
        =&\sum_{\alpha=1}^L\sum_{\beta\neq\alpha}\sum_{k=1}^{M_\alpha}\sum_{\ell=1}^{M_\beta}p_\lambda(k|\alpha)p_\lambda(\ell|\beta)
        fa_\alpha a_\beta
        \\&+\sum_{\alpha=1}^L\sum_{k=1}^{M_\alpha}\sum_{\ell\neq k}p_\lambda(k|\alpha)p_\lambda(\ell|\alpha)
        c_\alpha a_\alpha^2
        \\&+\sum_{\alpha=1}^L\sum_{k=1}^{M_\alpha}p_\lambda(k|\alpha)^2b_\alpha a_\alpha^2\\
        =&f\sum_{\alpha=1}^L\sum_{\beta\neq\alpha}a_\alpha a_\beta
        \\&+\sum_{\alpha=1}^L\sum_{k=1}^{M_\alpha}p_\lambda(k|\alpha) \Big[1-p_\lambda(k|\alpha)\Big]c_\alpha a_\alpha^2
        \\&+\sum_{\alpha=1}^L b_\alpha a_\alpha^2\sum_{k=1}^{M_\alpha}p_\lambda(k|\alpha)^2\\
        =&fA^2+\sum_{\alpha=1}^L a_\alpha^2(c_\alpha-f)+\sum_{\alpha=1}^L a_\alpha^2
        \mathcal{C}_\alpha(b_\alpha-c_\alpha),
    \end{split}
    \]
    where in the last line we introduced the index of coincidence
    \begin{equation}
\mathcal{C}_\alpha\equiv\sum_{k=1}^{M_\alpha}p_\lambda(k|\alpha)^2.
    \end{equation}
 From the properties of the generalized equiangular measurement parameters, we know that \cite{GEAM}
    \begin{equation}\label{cf}
        c_\alpha-f=-\frac{db_\alpha-1}{d(M_\alpha-1)}=-\frac{b_\alpha-c_\alpha}{M_\alpha}.
    \end{equation}
    This allows us to further simplify the formula for $\Tr(Y_\lambda^2)$ to
    \begin{equation}
    \begin{split}
        \Tr(Y_\lambda^2)=A^2f+\sum_{\alpha=1}^La_\alpha^2(f-c_\alpha)(M_\alpha\mathcal{C}_\alpha-1).
    \end{split}
    \end{equation}
Therefore, for $X=Y_\lambda$ in Lemma 1, we identify $m=A/d$ and
\[
        \begin{split}
        s^2=&\frac{1}{d-1}(\Tr Y^2-dm^2)
        \\=&\frac{1}{d-1}\left[A^2f+\sum_{\alpha=1}^La_\alpha^2(f-c_\alpha)(M_\alpha\mathcal{C}_\alpha-1)-\frac{A^2}{d}\right]\\
        =&\frac{1}{d-1}\sum_{\alpha=1}^La_\alpha^2(f-c_\alpha)
        (M_\alpha\mathcal{C}_\alpha-1).
        \end{split}
\]
Using the property $\mathcal{C}_\alpha\leq 1$, it becomes possible to bound $s^2$ with a quantity independent on the probability distributions $p_\lambda(k|\alpha)$;
    \begin{equation}
        \begin{split}
        s^2
        &\leq\frac{1}{d-1}\sum_{\alpha=1}^La_\alpha^2(f-c_\alpha)
        (M_\alpha-1)\\&=\frac{1}{d-1}\sum_{\alpha=1}^La_\alpha^2\left(b_\alpha-\frac 1d\right)\equiv\widetilde{s}^2,
        \end{split}
    \end{equation}
    where in the last equality we used eq. (\ref{cf}).
    Finally, after inputting our results into eq. (\ref{bounds}), we see that the maximal eigenvalue $\eta_{\max}(Y_\lambda)=\max_k\eta_k$ of $Y_\lambda$ is bounded by
\begin{equation}
    \xi_-\leq\eta_{\max}(Y_\lambda)\leq\xi_+,
\end{equation}
where
\begin{equation}\label{xip}
\begin{split}
    \xi_\pm&=m\pm\frac{d-1}{\sqrt{d}}\widetilde{s}\\&=\frac 1d \left[\sum_{\alpha=1}^La_\alpha\pm
    \sqrt{(d-1)\sum_{\alpha=1}^La_\alpha^2\left(db_\alpha-1\right)}\right].
\end{split}
\end{equation}
    
\section*{Appendix B: Asymptotic behavior of $\xi_+$ and unbounded violation}
 
Our goal is to show that there exist generalized equiangular measurements that lead to an unbounded steering violation. First, we assume that
\begin{equation}\label{db}
b_\alpha=\frac{z_\alpha}{d}.
\end{equation}
From the admissible condition on the parameters $b_\alpha$,
\[
\frac{1}{d} < b_\alpha \leq \frac{1}{d} \min\{d, M_\alpha\},
\]
it follows that $1<z_\alpha\leq\min\{d, M_\alpha\}$.
Substituting eq. (\ref{db}) and $a_\alpha=d\gamma_\alpha/M_\alpha$ into the general formula (\ref{xip}) for the LHS bound $\xi_+$ results in
\begin{equation}
\begin{split}
\xi_+=\sum_{\alpha=1}^L\frac{\gamma_\alpha}{M_\alpha}+
    \sqrt{\Delta(d)},
\end{split}
\end{equation}
where
\[
\Delta(d) := (d-1)\sum_{\alpha=1}^L\frac{\gamma_\alpha^2}{M_\alpha^2}(z_\alpha-1).
\]
Observe that $M_\alpha$ is at most a quadratic function in $d$ due to the constraint on the number of measurement elements \cite{GEAM}
\[
\sum_{\alpha=1}^NM_\alpha=d^2+N-1
\]
and the number of measurement settings $1\leq N\leq d^2-1$. Moreover, in order to sum up to the identity, $\gamma_\alpha$ have to be linear functions of $1/N$. Therefore, in the worst-case scenario, when $d\to\infty$, $\Delta(d)$ behaves like
\[
\Delta(d) \sim \frac{d^2}{NM_\alpha^2},
\]
where it was assumed that $L$ depends linearly on $N$. Therefore, for $\Delta(d)\to 0$, it is enough that $M_\alpha$ and $N$ are linear functions of the dimension $d$. In this case,
\[
\lim_{d\to\infty}\xi_+=\sum_{\alpha=1}^L\frac{\gamma_\alpha}{M_\alpha},
\]
which is always smaller than $1$ due to
\[
\sum_{\alpha=1}^L \frac{\gamma_\alpha}{M_\alpha}\leq
\sum_{\alpha=1}^N \frac{\gamma_\alpha}{M_\alpha}\leq
\sum_{\alpha=1}^N \frac{\gamma_\alpha}{2}=\frac 12.
\]
We use the fact that the smallest $M_\alpha=2$ and $\sum_{\alpha=1}^N\gamma_\alpha=1$.\\
This proves that, under mild assumptions on the number of measurement operators and $b_\alpha$, the LHS bound is always strictly less than $1$ for large enough dimensions.

\subsection*{Unbounded steering violation}


From definition, the quantum value of the steering functional is 
\[S_Q(F)=\sum_{\sigma\in\mathcal{Q}}|\<F,\sigma\>|,\]
where
  \[
    \begin{split}
    |\<F,\sigma\>|&=
\Tr\left(\sum_{\alpha=1}^L\sum_{k=1}^{M_\alpha}P_{\alpha,k}\sigma_{\alpha,k}\right)\\&\leq     \Tr\left(\sum_{\alpha=1}^L\sum_{k=1}^{M_\alpha}P_{\alpha,k}\sum_{\ell=1}^{M_\alpha}\sigma_{\alpha,\ell}\right)
\\&=\sum_{\alpha=1}^L\sum_{k=1}^{M_\alpha}\Tr\left(P_{\alpha,k}\rho_{\alpha}\right)=\sum_{\alpha=1}^L\gamma_\alpha\leq 1. 
    \end{split}
    \]
The first equality holds for the trivial assemblage $\sigma_{\alpha,k}=\mathbb{I}_d/d$ and the second for $L=N$. Therefore, $S_Q(F)=1$.
Now, the quantum violation is given by the formula
\[
V(F) = \frac{S_Q}{\xi_+}.
\]
If $\xi_+ < 1$, then $V(F) > 1$ and steering is detected. Furthermore, if $\xi_+ \to 0$,
\[
V(F) \to \infty,
\]
and therefore unbounded violation occurs.

\bibliographystyle{unsrt}
\bibliography{biblio_steering}

\end{document}